\RequirePackage{fixltx2e}
\documentclass[a4paper]{isp}
\usepackage[american]{babel}
\usepackage{graphicx}
\usepackage[section]{placeins}
\usepackage{paralist}
\usepackage[T1]{fontenc}
\usepackage{lmodern}
\usepackage{microtype}
\usepackage{amsmath}
\usepackage{booktabs}
\usepackage{url}
\ifnum\pdfoutput>0
\usepackage[
bookmarks=false,
breaklinks=true,
colorlinks=true,
linkcolor=black,
citecolor=black,
urlcolor=black,
pdfpagelayout=SinglePage
]{hyperref}
\usepackage[all]{hypcap}
\else
\usepackage{hyperref}
\fi
\usepackage[capitalise,nameinlink]{cleveref}
\crefname{section}{Sect.}{Sect.}
\Crefname{section}{Section}{Sections}
\crefname{figure}{Fig.}{Fig.}
\Crefname{figure}{Figure}{Figures}
\usepackage{xspace}
\newcommand{\eg}{it e.g. }
\newcommand{\ie}{it i.e. }
\newcommand{\aap}{    {\it Astron. Astrophys. }}

\newcommand{\apj}{    {\it Astrophys. J.  }}
\newcommand{\apjl}{    {\it  The Astrophysical Journal Letters}}

\newcommand{\roaj}{    {\it Romanian Astron. J. }}
\newcommand{\solphys}{{\it Solar Phys. }}

\begin{document}
\pdfgentounicode=1

%\title{Numerical simulations of the evolution of the complex AR12565 and AR12567}
\title{Numerical Simulations of the Evolution of Solar Active Regions: the Complex AR12565 and AR12567}
%If Title is too long, use \titlerunning
\titlerunning{Numerical Simulations of the Evolution of Solar Active Regions }

\author{Cristiana Dumitrache}
\institute{Astronomical Institute of Romanian Academy, Str. Cutitul de Argint 5, 040557 Bucharest, Romania\\
\email{crisd@aira.astro.ro}
}

\maketitle

\begin{abstract}
We have performed numerical magnetohydrodynamic (MHD) simulations of two closed active regions (AR). The input magnetic field values were the coronal magnetic field computed as extrapolation coronal from observations of the photospheric magnetic field.
The studied active regions, NOAA AR12565 and AR12567, were registered as different bipolar region. Our investigation, the 3D coronal extrapolations, as well as the numerical MHD experiments, revealed that actually they evolved together as a quadrupolar active region.
The second region emerged later under the loops system of AR12565 and separated from this one. A natural current sheet formed between then and it plays an important role in the explosive events (flares and coronal mass ejections) occurrence.

\end{abstract}

%%%%%%%%%%%%%%%%%%%%%%%%%%%%%%%%%%%%%%%%%%%%%%%%%%%%%%%%%%%%%%%%%%%%%%%%%%%%%%%
\section{Introduction}\label{sec:intro}
The active regions (ARs) represent major features of the solar activity. They are the typical manifestation of the bipolar or more complex magnetic field emerging from the convective zone. An active region is itself a complex entity involving different levels of the solar atmosphere, from the sub-photosphere to the corona.
Bumba and Howard \cite{Bumba1965} have provided an observational perspective of an active region development. They have highlighted the importance of the supergranular pattern in the formation of the sunspot groups, the base of the active regions.

Modeling such phenomena presume the development of non-linear processes and a multitude of developing scenarios, that's why the use of numerical simulations are the most common approaches.
Numerical MHD simulations, studying the formation and development of the active regions, approaches the rise of toroidal flux tubes from the bottom of the convection zone based on the thin flux tube approximation. There are simulations (\eg \cite{Roberts1978,Spruit1981,D’Silva1993,Caligari1995}) able to reproduce global properties of ARs such as their tilt angles, latitude of apparition, asymmetries between the leading and
following polarities. There are many other modern simulations and we mention the work of \cite{Cheung2010}, that emphasis the results from the interaction of the rising magnetic field with the turbulent convective flow during the rising process.
Fan and Gibson \cite{Fan2004} performed a 3D numerical simulation of a low-$\beta$ coronal magnetic field modeled as an arcade, where a twisted flux tube emerge into. They have found that the flux tube became kink-instable when it attainted a big enough twist. A current sheet with S-shape formed between the coronal and the emerging magnetic field.
Recently, \cite{Jiang2016} performed a MHD simulation of a nearly realistic simulation of a solar eruption from origin to onset, using observational data-driven model and a potential-field-source surface model for the coronal magnetic field extrapolation.
Inoue \cite{Inoue2016} summarized the current progress for the MHD  simulations using the observed magnetic field, by approaching both steps.

Our aim was to study the development of an active region starting from real magnetogram data as initial input for the magnetic field, and not with analytical expression as classical used.
Our goal is to emphasize the evolution of an active region, starting with the magnetic field data registered at a certain moment by HMI/SDO instrument, and to compare the obtained scenarios to the observed events.
A first attempt for such numerical MHD experiments was performed by \cite{dumi2015}.
This goal implied two stages of the work: (I) the extrapolation of the coronal magnetic field from the observed photospheric one, based on nonlinear force-free field approximation, and (II) the use of this magnetic field components as initial input for the MHD numerical simulations.
We have applied a nonlinear force-free field method to extrapolate the magnetic field, method transcribed in an IDL program by \cite{Lee2002}, to obtain the components of the magnetic field $Bx,By,Bz$. Two of them, $Bx,Bz$ became the initial magnetic field values for a 2D MHD code based on SHASTA method. The Alfven code was described by Wim Weber in his PhD thesis\cite{Weber1978}, and later intensively used by T.Forbes \cite{Forbes82}, and C.Dumitrache \cite{Dumitrache99}.

The present work intends to study an interesting complex of two bipolar neighbor active regions that developed together.
We have chosen the AR12565 and AR12567 that were observed on the Sun between 12 and 24 July 2016. The bipolar AR12565 raised the first on the visible side of the Sun, on 12 July 2016, and it was followed by AR12567, forming together a quadrupolar active region on 16 July 2016. They developed as a unique quadrupolar active region for few days.
Both active regions were very productive in flares and more mass ejecta.
We have observed these two regions in white light and $H_{alpha}$ wavelength at Bucharest Solar Station of the Astronomical Institute of Romanian Academy, where three solar flares were registered: on 18 July - a two ribbon flare, while on 21 and 22 July - compact flares \cite{Mouhamed2016}.

\section{Description of the method}
The computational method implies two main steps:
(I) the preparation of the data by performing 3D extrapolation of the coronal magnetic field lines and so getting the values of the magnetic field components $Bx, By, Bz$ at that time moment,
and (II) the MHD numerical simulations in the coronal conditions, using a 2D code and the $Bx, Bz$ components.

\begin{figure}
  \centering
  \includegraphics[width=0.8\textwidth]{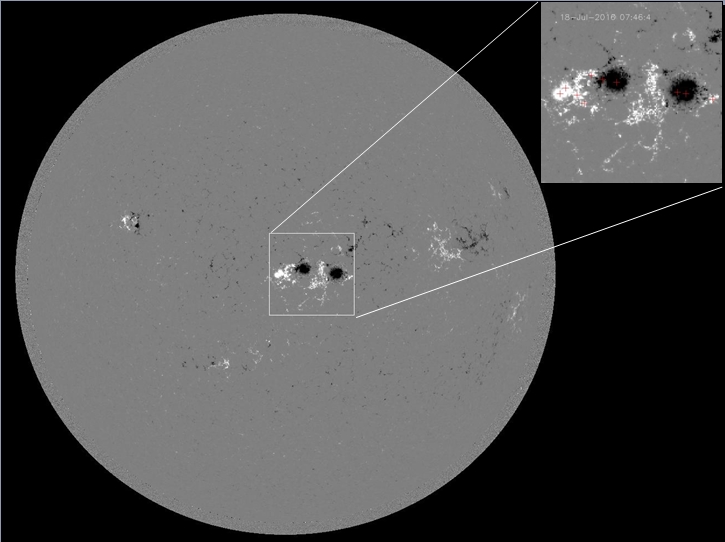}
  \caption{The full-disk HMI/SDO magnetogram and the active region zone cutting up. }\label{f1}
\end{figure}

(I) We have used the magnetograms measured by the HMI instrument onboard of the SDO spacecraft.
First, the active regions zone was selected and cut off from the magnetogram in order to be analysed.
Fig. \ref{f1} displays the magnetogram and the active regions zones, where the little red crosses indicate the ten dipoles positions used in the extrapolation computation at that moment.

The positions of the five pairs of magnetic dipoles were detected in Cartesian coordinates from the magnetogram, by choosing the local extrema for each polarity by Powell method. Finding the locations of the dipoles from the magnetogram involves two steps:
finding the \emph{XY} coordinates of the major or dominate dipoles and than finding the z coordinates.

The IDL code \cite{Lee2002} considers the active region as a magnetic dipole defined by
\begin{eqnarray*}
Bx &=&B0\cdot \frac{3xz}{r^{5}} \\
By &=&B0\cdot \frac{3yz}{r^{5}} \\
Bz &=&-B0\frac{(1-3\cdot \frac{z^{2}}{r^{2}})}{r^{3}}
\end{eqnarray*}%
where $r=\sqrt{x^{2}+y^{2}+z^{2}}$ and $B0$ scales the strength of the magnetic field.
We have modified this code to be adapted to our needs. The numerical resulted components of the magnetic field were used to the next step (II).

%%%%%%%%%%%%%%%%%%%%%%%%%%%%%%%%%%%%%%%%%%%%
(II) The second step was performed using a two dimensions Fortran code for the integration of the MHD equations. The Alfven code was described by \cite{Weber1978} and extensively used by T.Forbes, and later by C.Dumitrache (\cite{Forbes82}, \cite{Dumitrache99}). This code is based on SHASTA (SHarp and Smooth Transport Algorithm) method that was developed by Boris and Book \cite{BorisBook1973}, \cite{BorisBook1975}. It is a flux-corrected transport (FCT) algorithm type that was constructed to solve the radiative, diffusive MHD equations by a predictor-corrector numerical scheme of integration. SHASTA uses high-diffusion monotone first-order schemes and low-diffusion second-order ones.
We point out that the Alfven code computations published in the past were performed on a mesh 49 x 97. For big mesh, the code becomes instable.
After a long experience of working with this code, we considered recently a mesh of 100 x 100, taking care for the numerical step of integration.

The MHD equations (1)--(6) are integrated on a mesh $100 \times 100$ points:

\begin{equation}
\dfrac{\partial \rho }{\partial t}+\overrightarrow{\nabla }(\rho \cdot
\overrightarrow{v})=0\
\end{equation}

\begin{equation}
\rho [\dfrac{\partial \overrightarrow{v}}{\partial t}+(\overrightarrow{v}%
\cdot \overrightarrow{\nabla })\overrightarrow{v}]=-\overrightarrow{\nabla }%
(p)+(\overrightarrow{B}\cdot \overrightarrow{\nabla })\cdot \overrightarrow{B%
}-\overrightarrow{\nabla }\cdot (\dfrac{B^{2}}{2})+\rho \cdot
\overrightarrow{g}\
\end{equation}

\begin{equation}
\dfrac{\partial \overrightarrow{B}}{\partial t}=\overrightarrow{\nabla }%
\times (\overrightarrow{v}\times \overrightarrow{B})+\eta \overrightarrow{%
\nabla }^{2}\cdot \overrightarrow{B}
\end{equation}

\begin{equation}
\dfrac{\rho ^{\gamma }}{\gamma -1}\dfrac{d}{dt}(\dfrac{p}{\rho ^{\gamma }})=-%
\overrightarrow{\nabla }(k\overrightarrow{\nabla }T)-\rho
^{2}Q(T)+j^{2}/\sigma +h\rho
\end{equation}

\begin{equation}
p=\rho T
\end{equation}
%%%%%%%%%%%%%%%%%%%
where $\overrightarrow{B},\overrightarrow{v},\rho ,p,T$ are dimensionless variables defined by

\begin{eqnarray}
\rho &=&\rho ^{\prime }/10^{9\text{ }}  \label{7} \\
T &=&T^{\prime }/10^{6}  \nonumber \\
p &=&4\pi p^{\prime }/Bo^{2}  \nonumber \\
\overrightarrow{v} &=&\overrightarrow{v}^{\prime }/v_{a}  \nonumber \\
\overrightarrow{B} &=&\overrightarrow{B}^{\mathbf{\prime }}/Bo  \nonumber \\
t &=&\dfrac{t^{^{\prime }}\cdot v_{a}}{w^{\prime }}\text{ }  \nonumber
\end{eqnarray}
and $\overrightarrow{B^{\prime }},\overrightarrow{v^{\prime }},\rho ^{\prime}, p^{\prime },T^{\prime },t^{\prime }$ are respectively the dimensional
magnetic field intensity, plasma flow velocity, density, gas pressure,
temperature and time. $v_{sc}$ denotes the sound velocity, $\gamma =\frac{5}{3%
}$ is the ratio of the specific heats and $\eta $ is the dimensionless
resistivity. In term of $\eta $, the dimensionless magnetic Reynolds number
is $R_{m}=\eta ^{-1}=v_{a}w^{\prime }/\eta ^{\prime }$, with Alfven speed
velocity $v_{a}=Bo/(4\pi \rho _{o}^{\prime })^{1/2}$; here $Bo=1.57$ G and $%
\rho _{o}^{\prime }=10^{9}$ cm$^{-3}$, where $\eta ^{\prime }$ is the
resistivity and $w^{\prime }$ is the characteristic scale-length of the
initial conditions (\emph{i.e.} the width of the initial current sheet). The
dimensionless time, $t$, and spatial coordinates, $x$ and $z$, are related
to their dimensional counterparts, $x^{\prime },z^{\prime },t^{\prime }$, by
\[
x=x^{\prime }/R_{\odot },\ z=w/z_{o}^{\prime },\ t=t^{\prime }/t_{a}
\]
where $z_{o}^{\prime }=2w^{\prime }$ is the computational box, $%
t_{a}=w^{\prime }/v_{a}$ is the Alfv\'{e}n scale-time and $R_{\odot }$ is
the solar radius.

The boundary conditions used for the computational grid are:

-at top (x=1)
\begin{equation}
\frac{\partial B_{z}}{\partial x}=\frac{\partial B_{z}}{\partial x}=0
\end{equation}

\begin{equation}
\frac{\partial B_{z}}{\partial x}=-\frac{\partial B_{x}}{\partial z}
\label{9}
\end{equation}

- at right (z=1)
\begin{equation}
\frac{\partial B_{x}}{\partial z}=\frac{\partial B_{z}}{\partial z}=0
\label{10}
\end{equation}
\begin{equation}
\frac{\partial B_{z}}{\partial z}=-\frac{\partial B_{x}}{\partial x}
\label{11}
\end{equation}

- at left - the symmetry axis (z=0)
\begin{equation}
\frac{\partial B_{x}}{\partial z}=B_{z}=0  \label{12}
\end{equation}

- at bottom (x=0)
\begin{equation}
\frac{\partial B_{x}}{\partial x}=\frac{\partial B_{z}}{\partial x}=0
\label{13}
\end{equation}

In the present paper, we have performed both steps of calculations for more magnetograms, registered at different time moments, for $\beta=0.1$ and one solar radius.
The run of the program for a certain magnetogram was stopped when a stable state is attainted and the topology of the magnetic field, velocity field and plasma density rest unchanged. Usually, we have obtained development of a configuration for few seconds.
The use of a 2D code is enough for our purpose, for the moment. The results of the simulations are only indicative for the plasma and magnetic field behaviour, but gave unrealistic values for the plasma parameters.

\section{The results}
Our simulations were focused on the period when both region are visible on the Sun, \ie between 16 and 22 July.
We have analyzed and started with the magnetograms registered on the following dates and times format $yyyymmdd \_ hhmmss$.
Each figure displays the results of the 3D coronal magnetic field extrapolations at a time moment, magnetic field ($Bx$ and $Bz$ components) that is the starting point for the simulations. The main stages of the numerical results are plotted too, where the elapsed time in seconds is noted at the up left corner of each cartoon.

The cartoons with the results plot the dimensionless quantities: the magnetic field lines, the velocity field and the plasma density (filed plots), on a grid of $200 \times 100$, \emph{ie} the symmetry axis is at the 10 division on the graphic abscise. In the plots, the Ox axis from the simulations code, along that the gravity acts, becomes Oz, \ie vertical one.
Each figure displays  the bar code of the colors  used for the density for the filled contours. The bar code labels represent the values of the dimensionless density ($\rho$), the dimension coronal densities being $\rho^{`} = \rho \cdot 10^9$  part./ cm$^{-3}$.

\begin{figure}[!htb]
\begin{center}
\includegraphics[width=5cm,height=!]{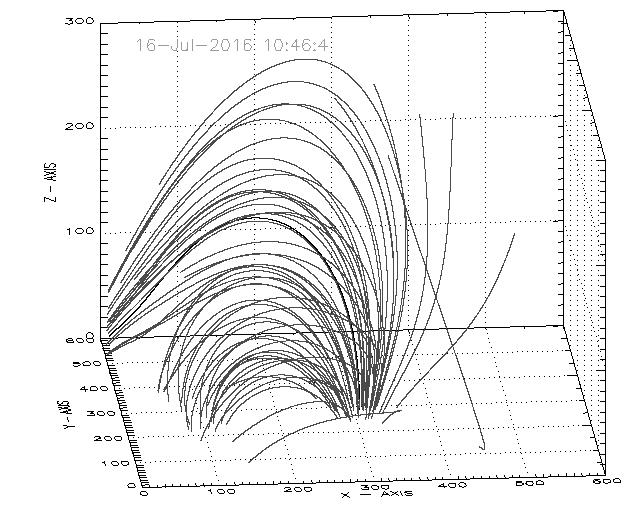}\\
\includegraphics[width=3.5cm,height=4cm]{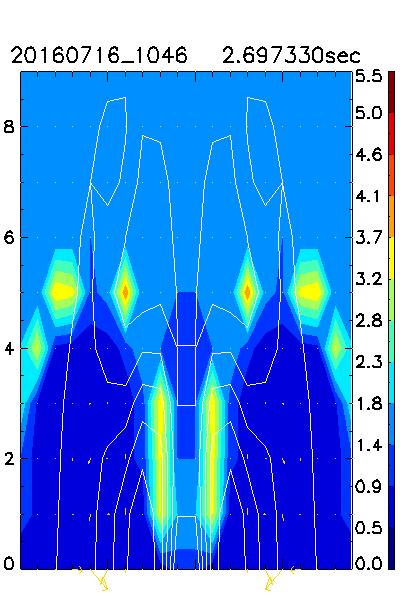}
\includegraphics[width=3.5cm,height=4cm]{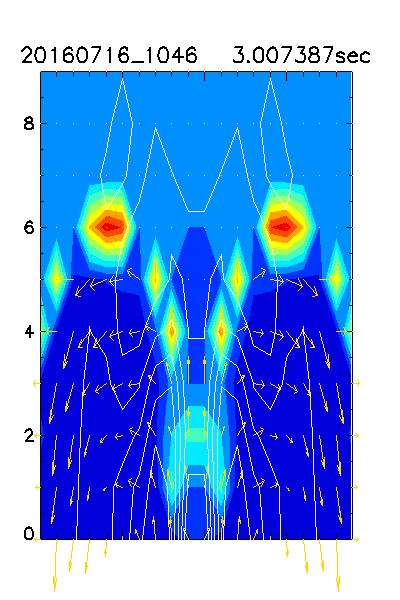}
\includegraphics[width=3.5cm,height=4cm]{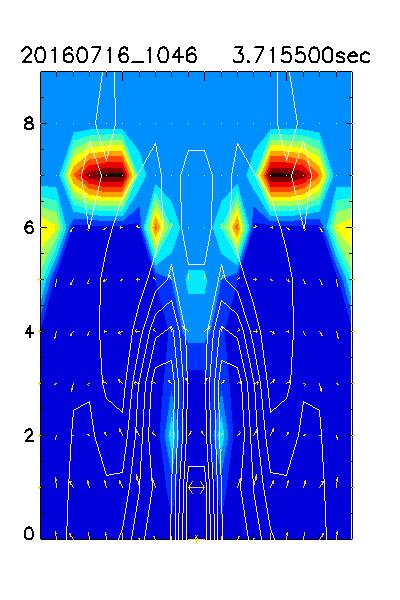}\\
\includegraphics[width=3.5cm,height=4cm]{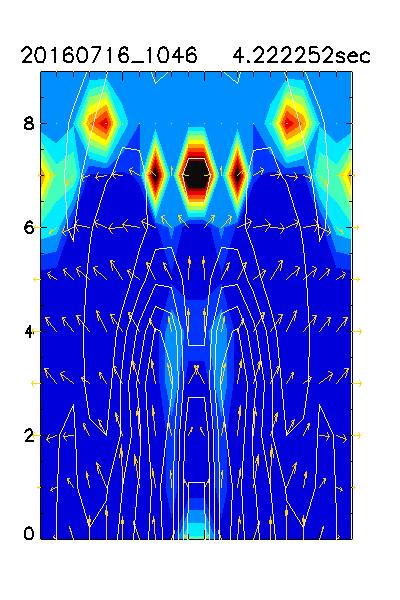}
\includegraphics[width=3.5cm,height=4cm]{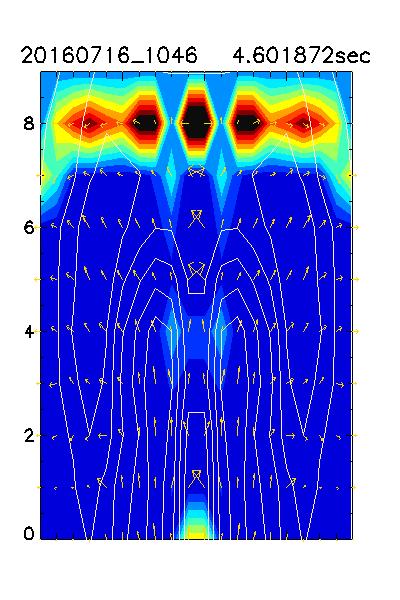}
\includegraphics[width=3.5cm,height=4cm]{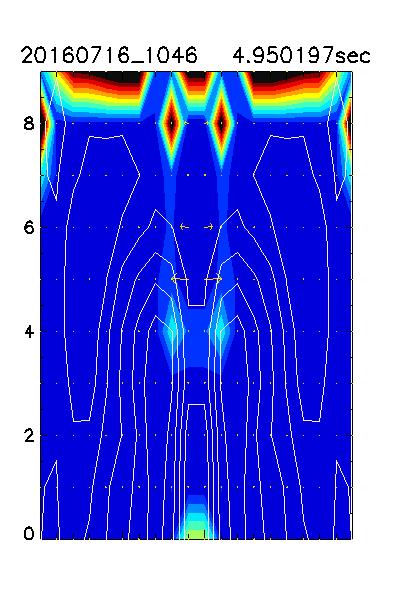}
  \caption{The simulation results for 16 July, 10:46 UT: the density (filled contours), magnetic field lines (plot contours) and velocity fields. The colour bar quantifies the dimensionless density.}
  \label{f16a}
\end{center}
\end{figure}

\begin{figure}[!htb]
\begin{center}
\includegraphics[width=5cm,height=!]{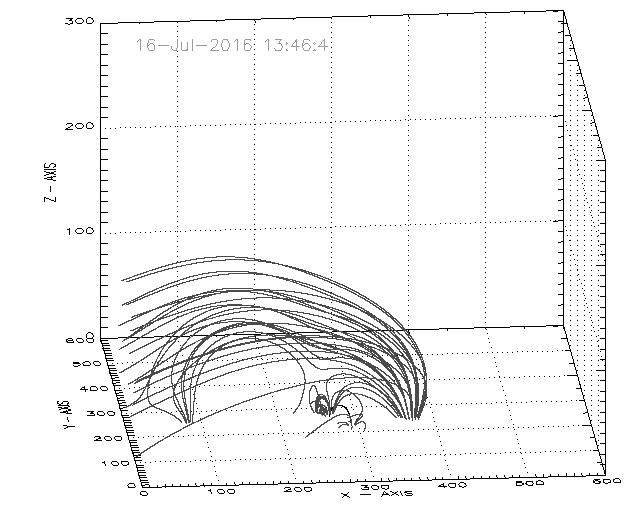}\\
\includegraphics[width=3.5cm,height=4cm]{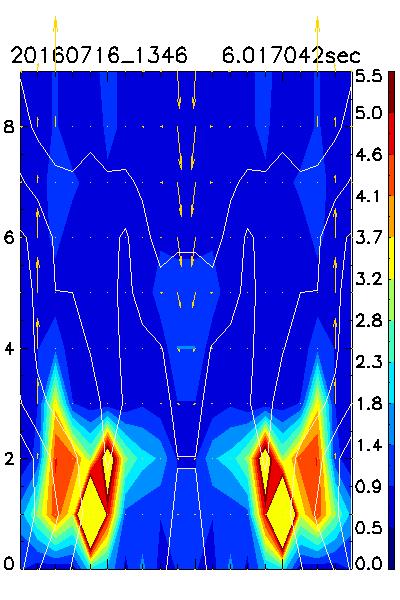}
\includegraphics[width=3.5cm,height=4cm]{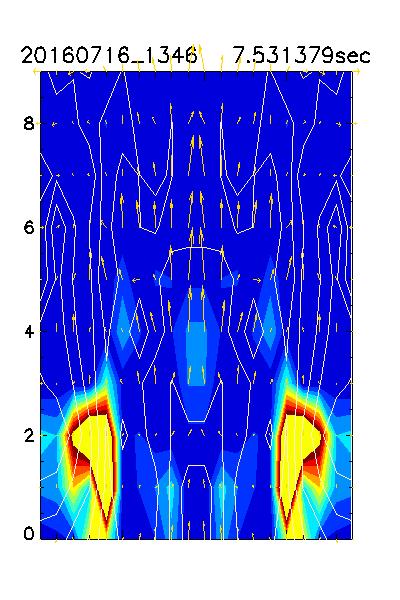}
\includegraphics[width=3.5cm,height=4cm]{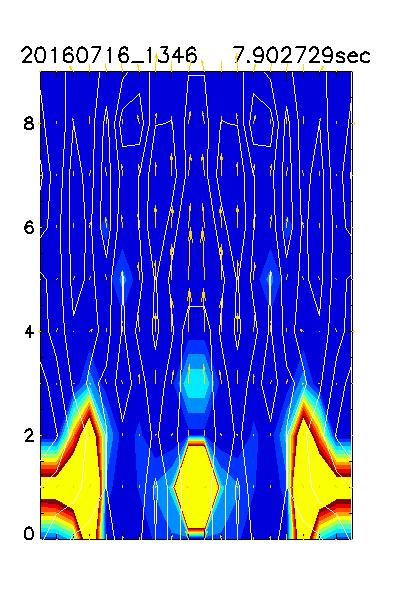}\\
\includegraphics[width=3.5cm,height=4cm]{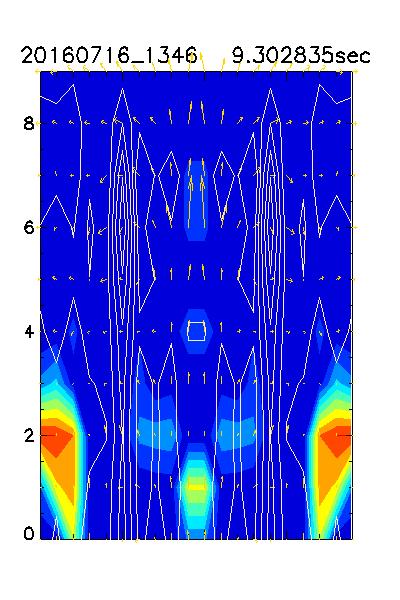}
\includegraphics[width=3.5cm,height=4cm]{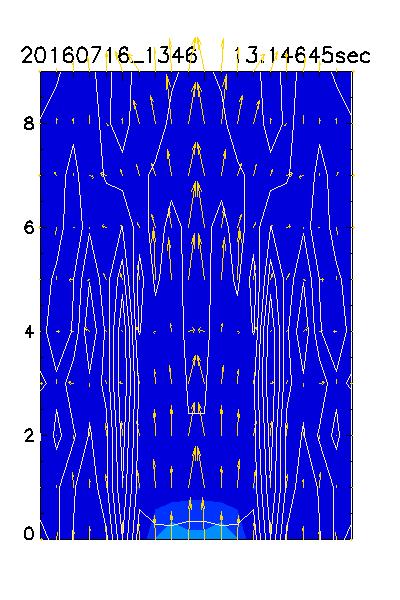}
\includegraphics[width=3.5cm,height=4cm]{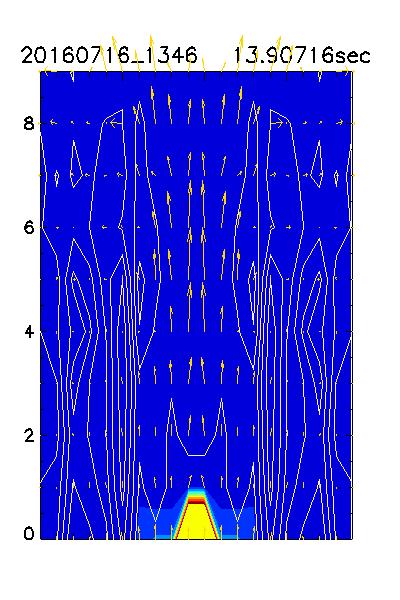}
  \caption{The simulation results for 16 July, 13:46 UT: the density (filled contours), magnetic field lines (plot contours) and velocity fields. The colour bar quantifies the dimensionless density.}
  \label{f16b}
\end{center}
\end{figure}

The first magnetogram contents the AR12565 field lines, but the AR12567 emerged during 16 July 2016. Next days, during their evolution, these two active regions acted either as one or as two entities. They produced many flares and suffered many magnetic reconnections. The open magnetic field lines in the figures are marking of eruptive events.
Fig. \ref{f16a} and \ref{f16b} display the evolution of the complex composed by NOAA AR12565 and AR12567 during the day when the second active region appeared very closed to AR12565.
As fig. \ref{f16a} indicates, the first cartoon catches the flux emergence. The open magnetic field lines indicates the place of emergence and they are bought together as in a single root.
The simulations (\ref{f16a}) show the plasmoids emergence from an unique root, with the development of plasma flows in divergent directions. Two loop systems formed later, starting with the third cartoon in the figures. At the end of the simulation we remark a flare resulted from all the magnetic reconnections in zone. To note that a flare can be also seen on the AIA/SDO coronal observations in the 131${\AA}$ wavelength (please see \cite{aia}).
In fig. \ref{f16b}, we see in the 3D extrapolations the completed apparition of the second active region and the well defined loop systems of both regions, with also loops underlying both of them. The simulations also indicate the separation of a bipolar active region in two parts, but with plasma emergence in the current sheet appeared between the two loop systems. Interesting fact is that downward flows can be view in the first stage of evolution, flows that probably pushed and separated the active regions, before the loops form. After that, upward flows indicate continuum flux emergence.

%%%%%%%%%%%%%%%%%%%%%

\begin{figure}[!htb]
\begin{center}
 \includegraphics[width=5cm,height=!]{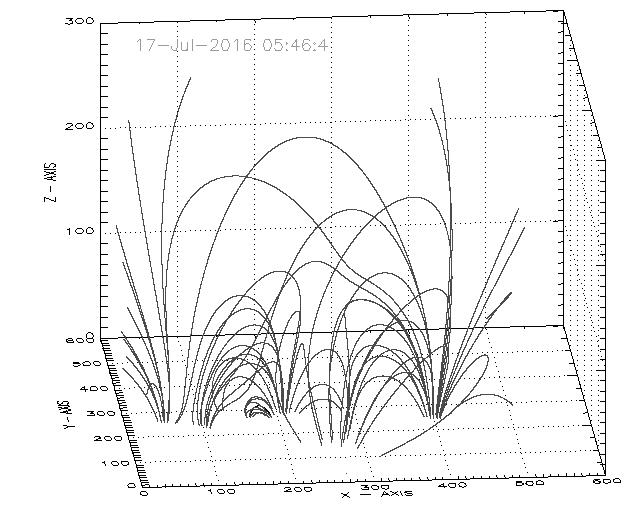}\\
\includegraphics[width=3.5cm,height=4cm]{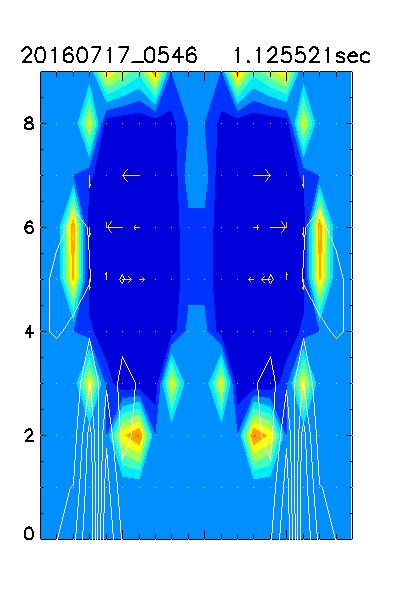}
\includegraphics[width=3.5cm,height=4cm]{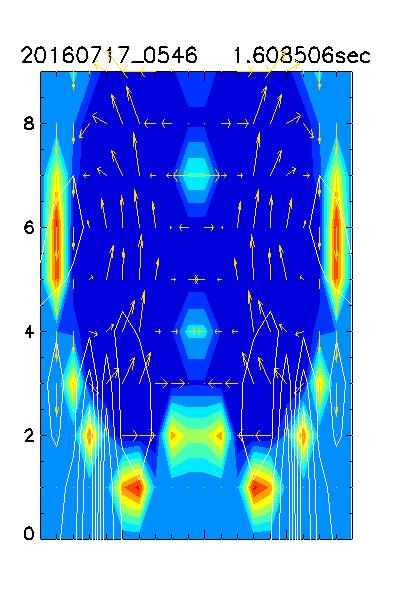}
\includegraphics[width=3.5cm,height=4cm]{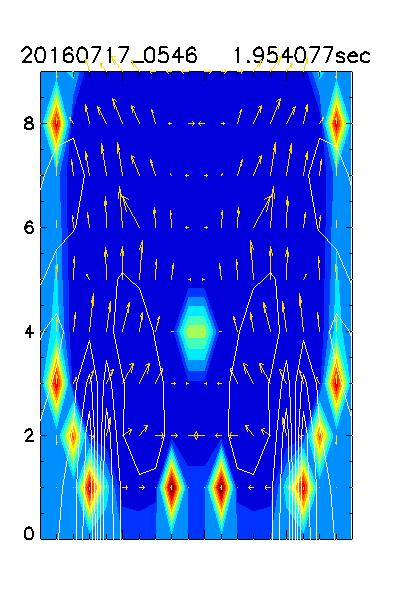}\\
\includegraphics[width=3.5cm,height=4cm]{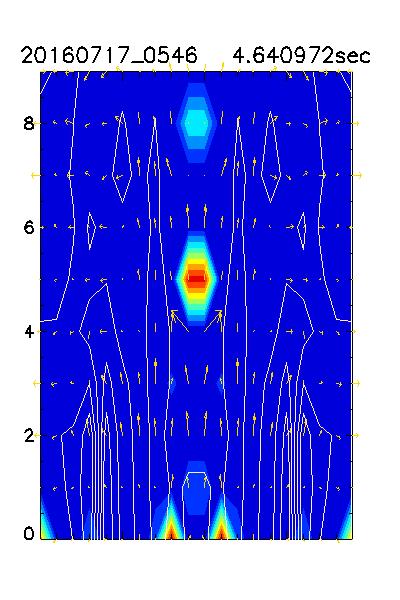}
\includegraphics[width=3.5cm,height=4cm]{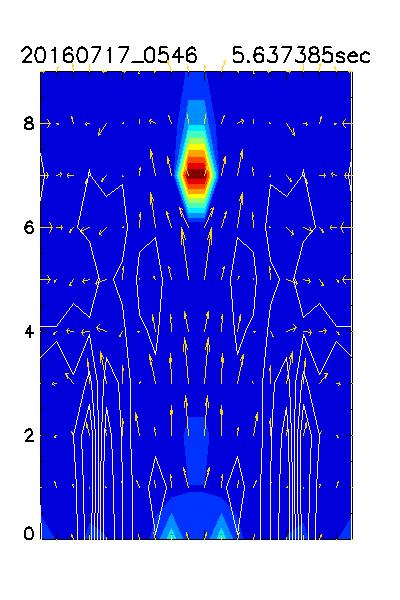}
\includegraphics[width=3.5cm,height=4cm]{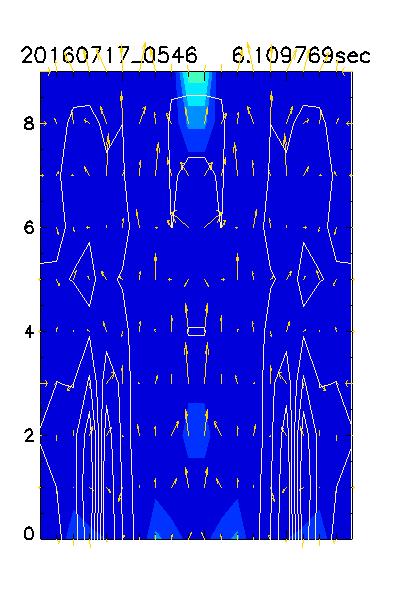}
  \caption{The simulation results for 17 July, 05:46 UT: the density (filled contours), magnetic field lines (plot contours) and velocity fields. The colour bar quantifies the dimensionless density.}
  \label{f17}
\end{center}
\end{figure}

\begin{figure}[!htb]
\begin{center}
\includegraphics[width=5cm,height=!]{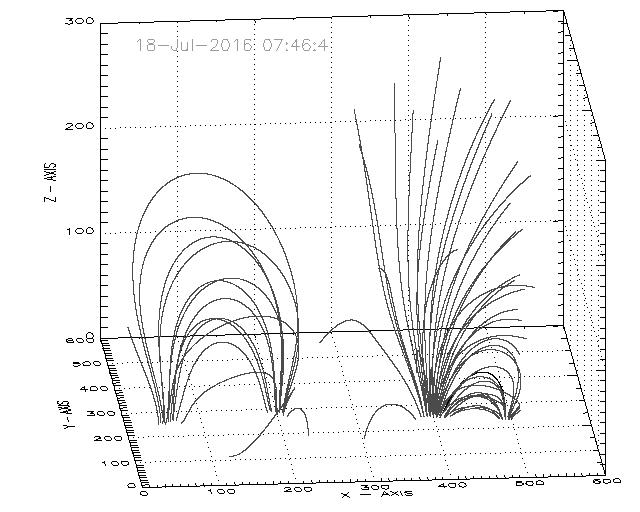}\\
\includegraphics[width=3.5cm,height=4cm]{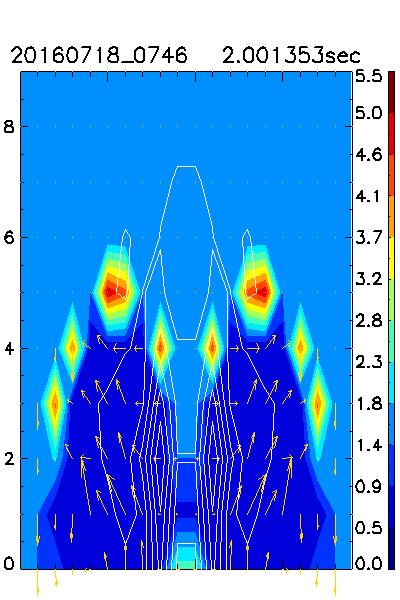}
\includegraphics[width=3.5cm,height=4cm]{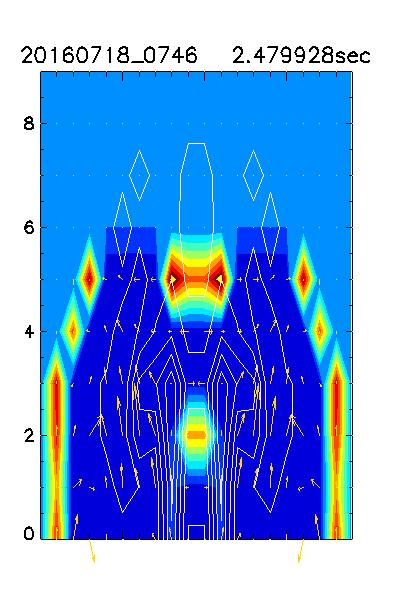}
\includegraphics[width=3.5cm,height=4cm]{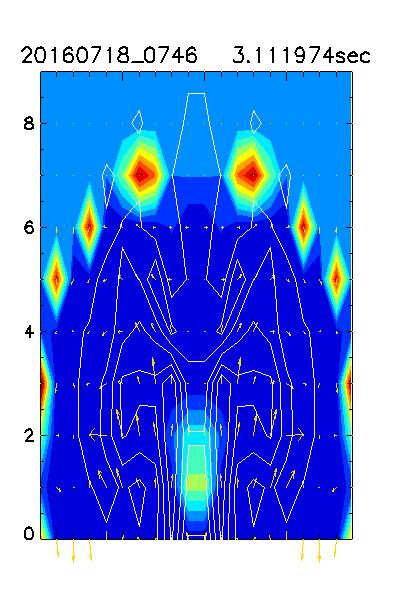}
\includegraphics[width=3.5cm,height=4cm]{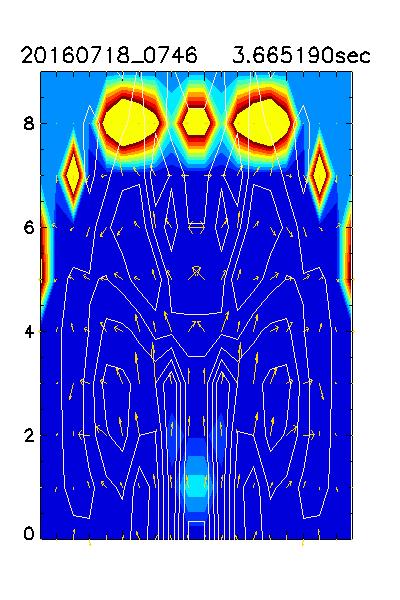}
  \caption{The simulation results for 18 July, 07:46 UT: the density (filled contours), magnetic field lines (plot contours) and velocity fields. The colour bar quantifies the dimensionless density.}
  \label{f18a}
\end{center}
\end{figure}

\begin{figure}[!htb]
\begin{center}
\includegraphics[width=5cm,height=!]{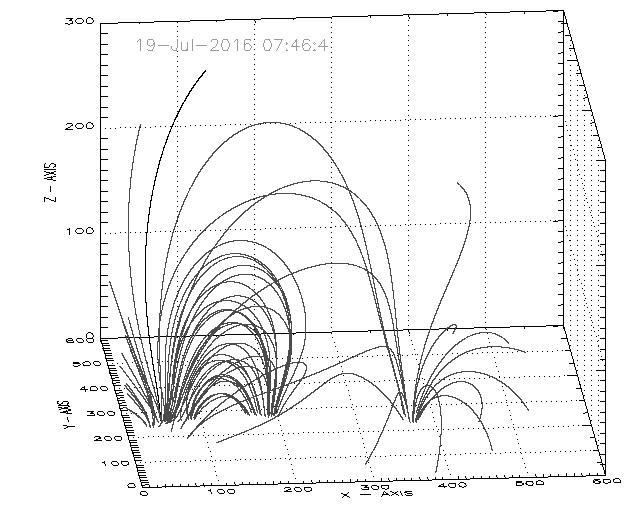}\\
\includegraphics[width=3.5cm,height=4cm]{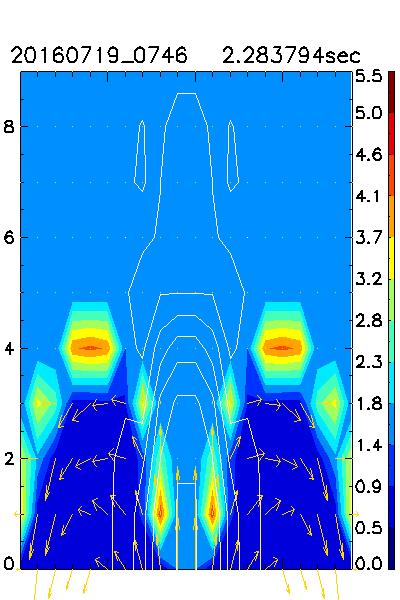}
\includegraphics[width=3.5cm,height=4cm]{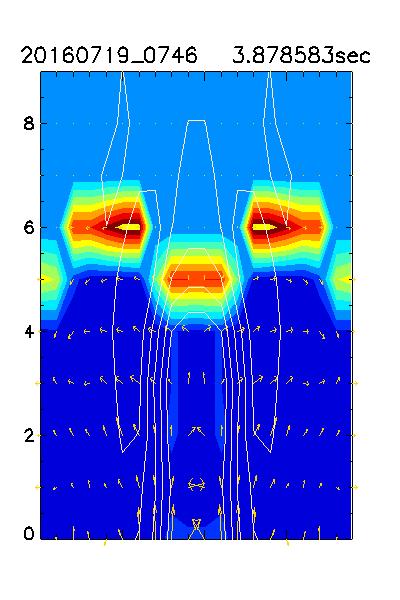}
\includegraphics[width=3.5cm,height=4cm]{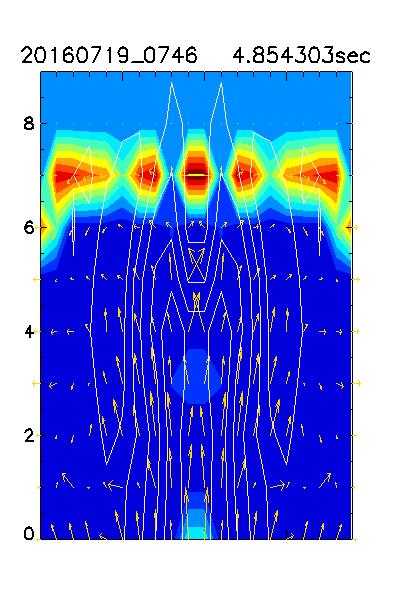}\\
\includegraphics[width=3.5cm,height=4cm]{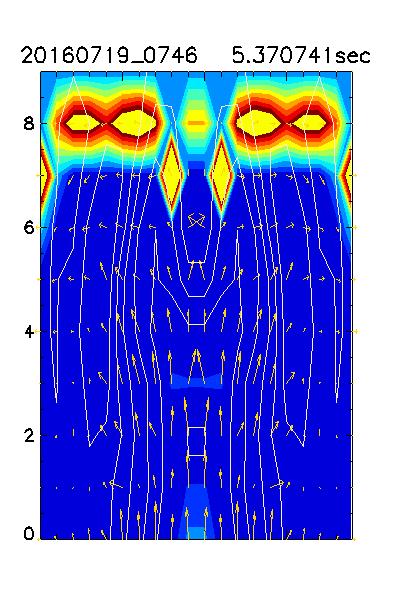}
\includegraphics[width=3.5cm,height=4cm]{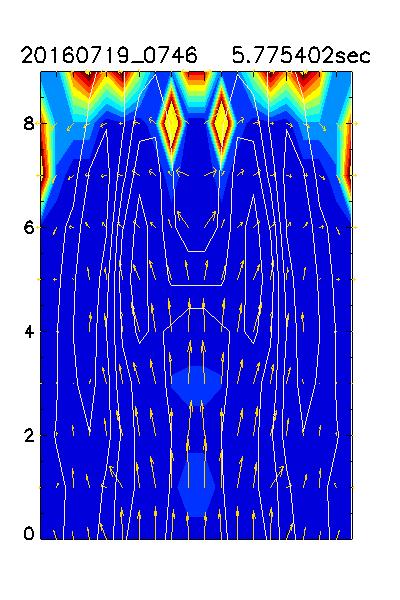}
  \caption{The simulation results for 19 July, 07:46 UT: the density (filled contours), magnetic field lines (plot contours) and velocity fields. The colour bar quantifies the dimensionless density.}
  \label{f19}
\end{center}
\end{figure}

\begin{figure}[!htb]
\begin{center}
\includegraphics[width=5cm,height=!]{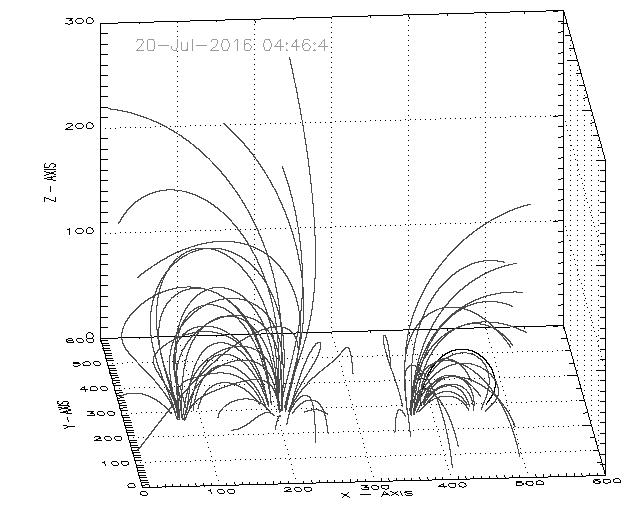}\\
\includegraphics[width=3.5cm,height=4cm]{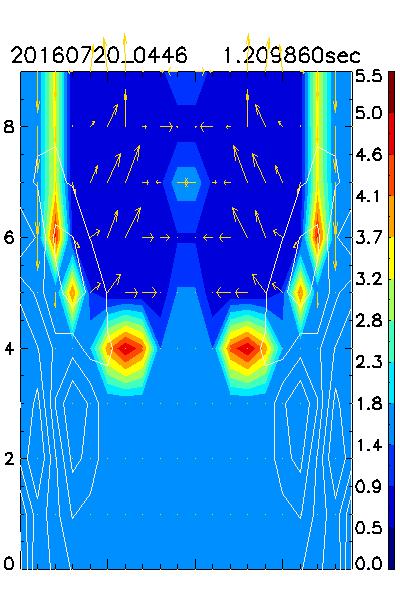}
\includegraphics[width=3.5cm,height=4cm]{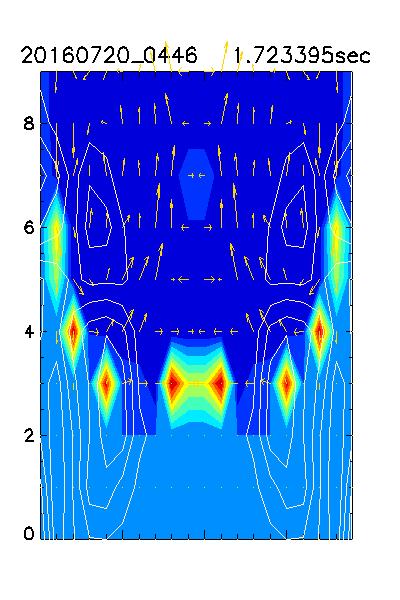}
\includegraphics[width=3.5cm,height=4cm]{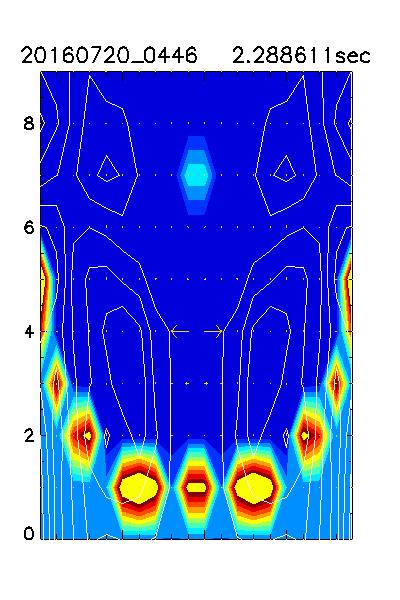}\\
\includegraphics[width=3.5cm,height=4cm]{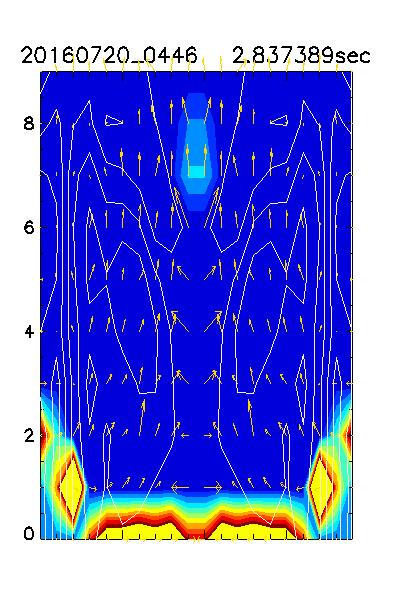}
\includegraphics[width=3.5cm,height=4cm]{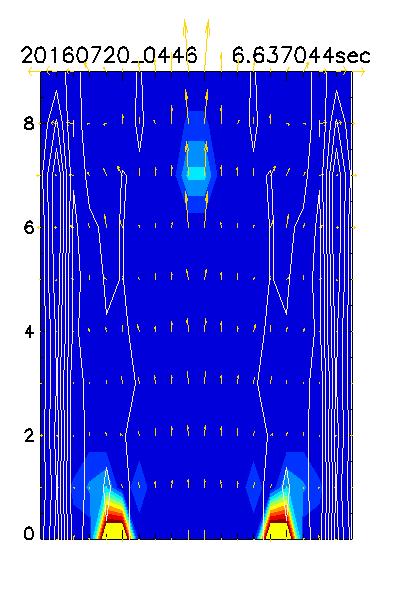}
\includegraphics[width=3.5cm,height=4cm]{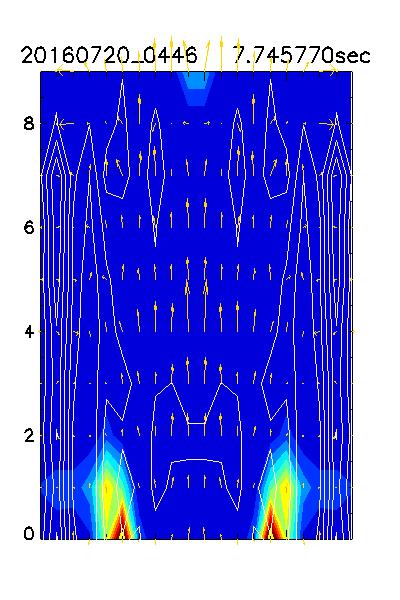}
  \caption{The simulation results for 20 July, 04:46 UT: the density (filled contours), magnetic field lines (plot contours) and velocity fields. The colour bar quantifies the dimensionless density.}
  \label{f20}
\end{center}
\end{figure}

\begin{figure}[!htb]
\begin{center}
\includegraphics[width=5cm,height=!]{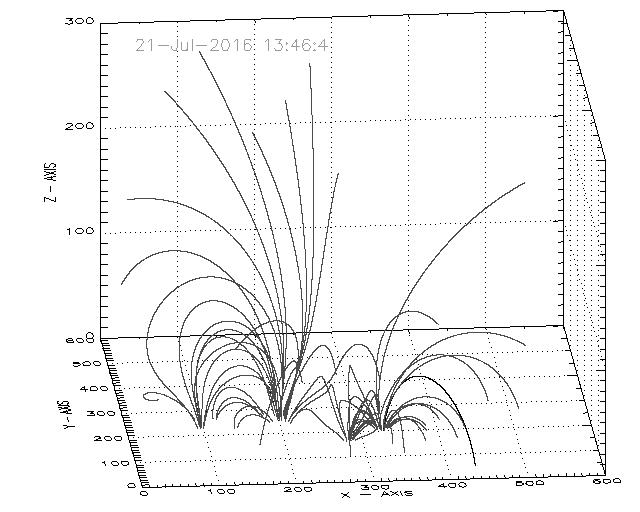}\\
\includegraphics[width=3.5cm,height=4cm]{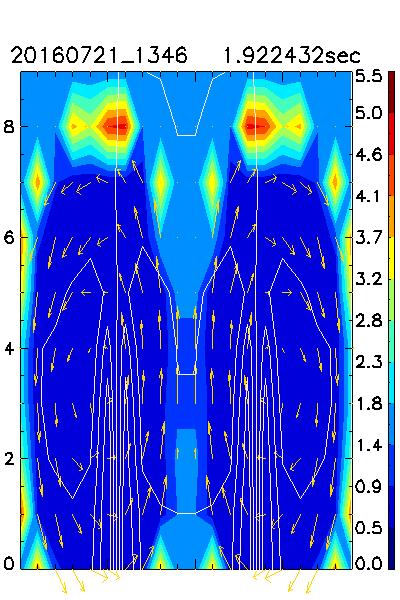}
\includegraphics[width=3.5cm,height=4cm]{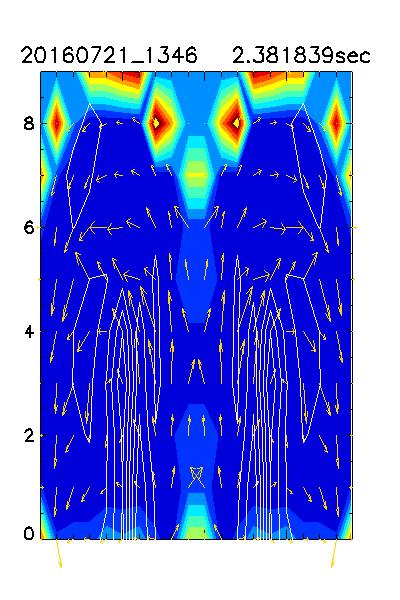}
\includegraphics[width=3.5cm,height=4cm]{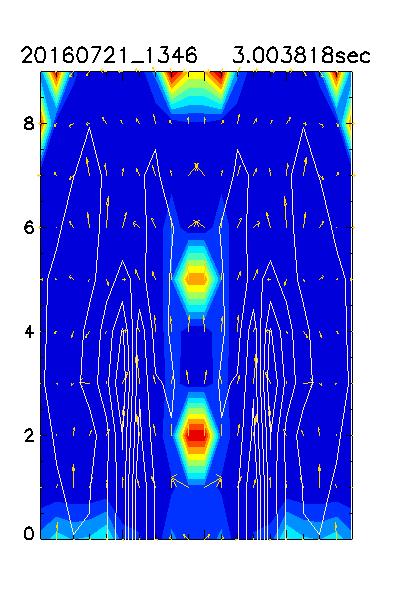}\\
\includegraphics[width=3.5cm,height=4cm]{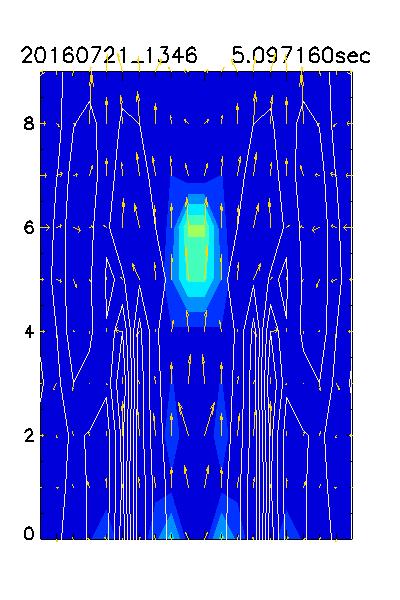}
\includegraphics[width=3.5cm,height=4cm]{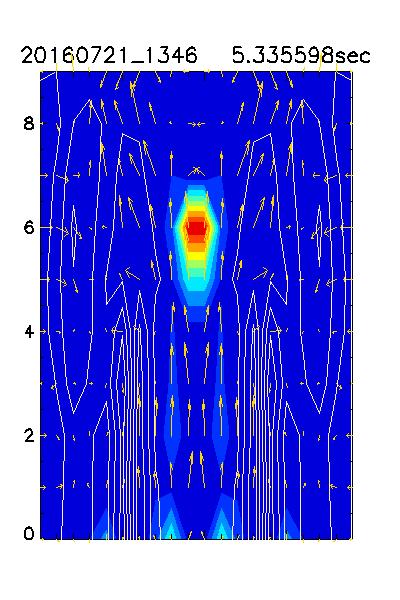}
\includegraphics[width=3.5cm,height=4cm]{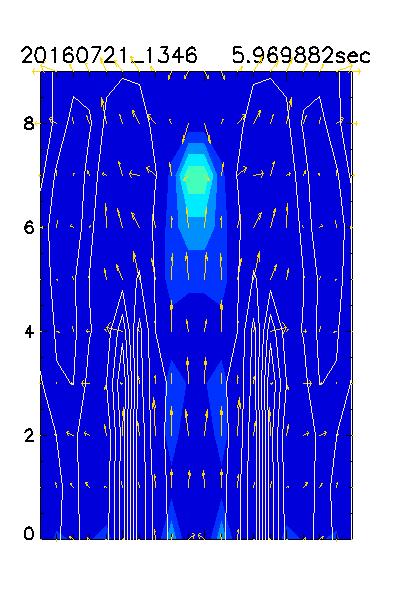}\\
\includegraphics[width=3.5cm,height=4cm]{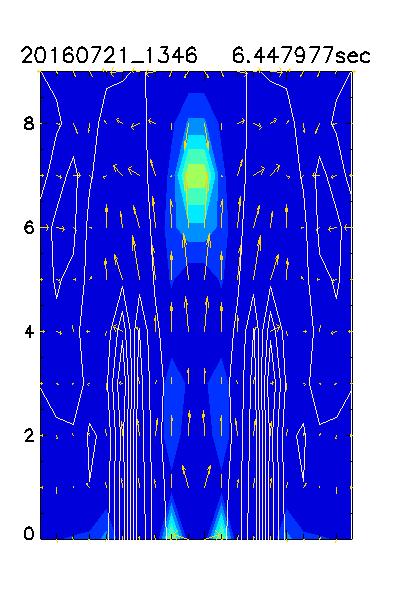}
\includegraphics[width=3.5cm,height=4cm]{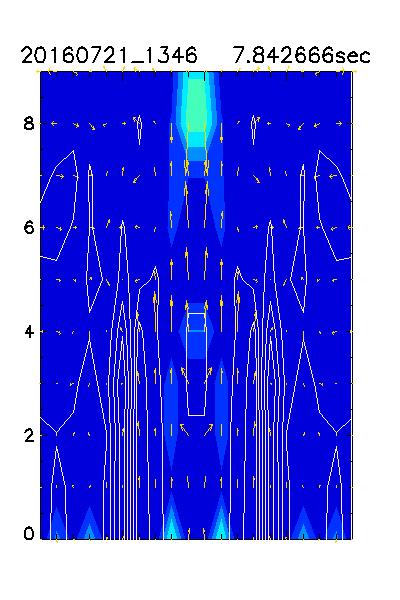}
\includegraphics[width=3.5cm,height=4cm]{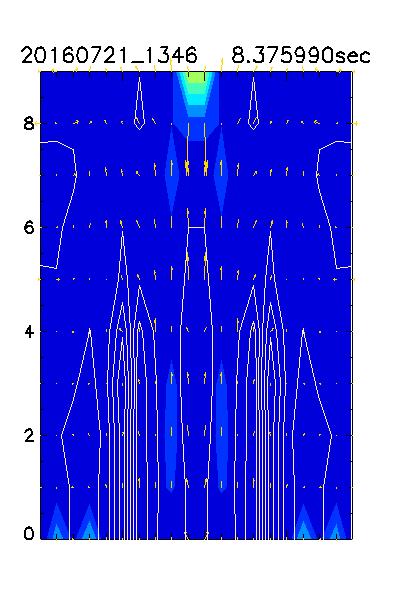}
  \caption{The simulation results for 21 July, 13:46 UT: the density (filled contours), magnetic field lines (plot contours) and velocity fields. The colour bar quantifies the dimensionless density.}
  \label{f21}
\end{center}
\end{figure}

\begin{figure}[!htb]
\begin{center}
\includegraphics[width=5cm,height=!]{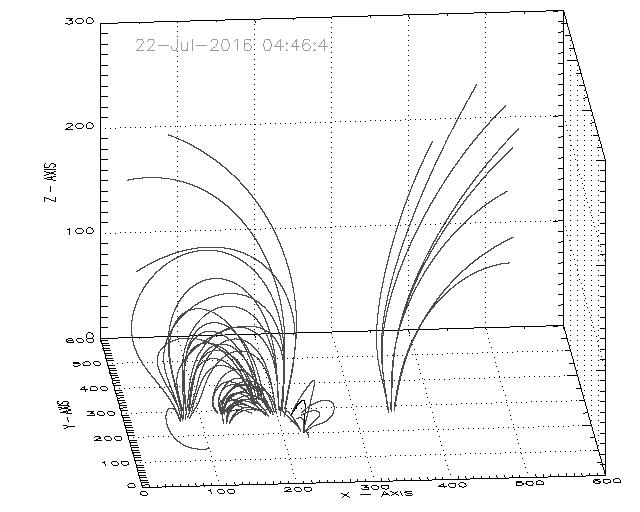}\\
\includegraphics[width=3.5cm,height=4cm]{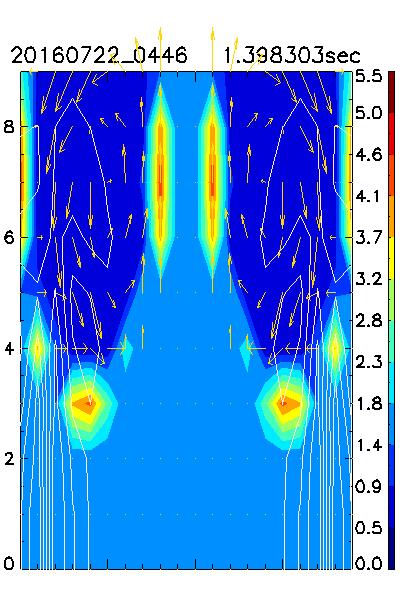}
\includegraphics[width=3.5cm,height=4cm]{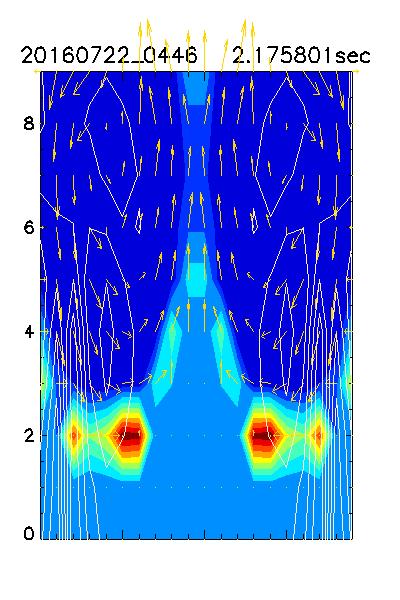}
\includegraphics[width=3.5cm,height=4cm]{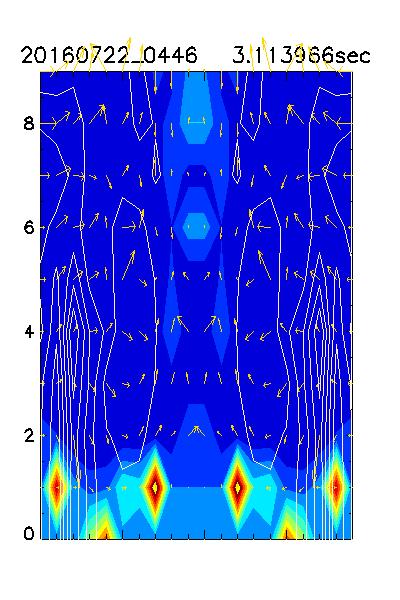}\\
\includegraphics[width=3.5cm,height=4cm]{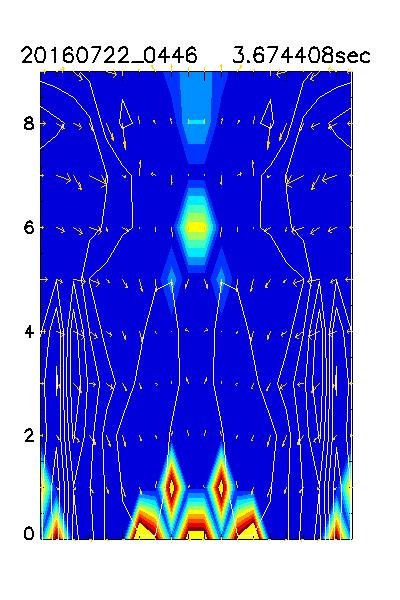}
\includegraphics[width=3.5cm,height=4cm]{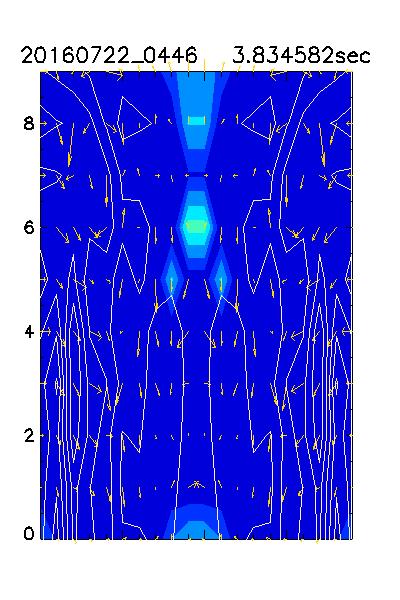}
\includegraphics[width=3.5cm,height=4cm]{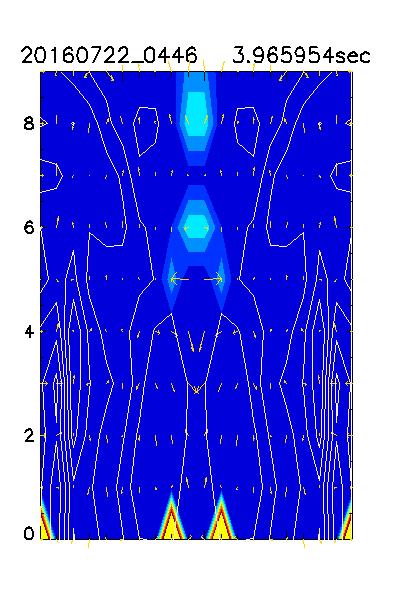}
  \caption{The simulation results for 22 July, 04:46 UT: the density (filled contours), magnetic field lines (plot contours) and velocity fields. The colour bar quantifies the dimensionless density.}
  \label{f22a}
\end{center}
\end{figure}

%%%%%%%%%%%%%%%%%
On 17 July the two loop systems representing the two active regions are well defined, with other loops underlying them in a unique quadrupolar system. In the region between them plasma insulated as the active regions push these plasmoids upward. A small coronal mass ejection (CME) can be observed in the middle region (\cite{cme17}). At that moment the active regions crossed the central meridian, but no other significative CME's source could be observed in the second dial on the solar disk \cite{solm17}.

On 18 July (fig. \ref{f18a}), both regions are separated and AR12565 displays open magnetic field lines in the 3D extrapolation. This magnetic configuration usually  indicates an explosive event that has produced or it follows to produce. The simulations show a continuum plasma and magnetic field emergence in zone between both active regions, and a two ribbon flare occurrence. The solar $H_{alpha}$ observations (\cite{Mouhamed2016}) testify this scenario: a two ribbon flare was observed as starting between AR12565 and AR12567 (at 07:54 UT), but extending at both active regions.
The simulations for 19 July (fig. \ref{f19}) repeated the same scenario. Observationally, few sub-flares and one more important flare were registered \cite{solmon}.

On 20 July (fig. \ref{f20}), AR12565 and AR12567 behave as two separate bipolar regions. Four sub-flares appeared (\cite{solmon}) and the numerical simulations indicate more features resembling with bright points.
On 21 July (fig. \ref{f21}) the active regions are linked again, but both of them display open magnetic field lines and indicate so explosive events to watch. The simulations revealed a flare followed by a CME. Observationally, more explosive events happen, one of them being observed also at Bucharest Observatory in $H_{alpha}$ line (\cite{Mouhamed2016}).

Fig. \ref{f22a} displays the results for 22 July, when AR12567 is more pronounced and AR12565 seems to be in the decaying phase, but with an open magnetic field configuration. The simulations revealed explosive events again and a mass ejecta. More flare were registered (\cite{solmon}) including one observed at Bucharest (\cite{Mouhamed2016}) starting with 06:38 UT. These simulations show again that both regions acted as single one since the main ejecta left the middle zone.
%%%%

\section{Discussions}
Far to be accomplished the goal of obtaining realistic values for the plasma parameters, as well as the aim of having MHD tool for explosive events forecast for the space weather purpose, we have performed a series of numerical magnetohydrodynamic (MHD) experiments starting with real magnetograms as input data for the magnetic field configuration.
These work involved two stages: the 3D coronal magnetic field extrapolation from HMI/SDO magnetograms and the integration of the MHD equations, by setting the bottom boundary of the model at the coronal base. The numerical scheme of the Fortran (Alfven) code used is based on a Flux-Corrected Transport algorithm (SHASTA method) and was applied in a Cartesian frame.
The results are interesting for the magnetic topology results, the velocity field and plasma density distribution, they giving a view of the scenarios of evolution for the studied active regions.
The comparison of these results to the observational data regarding the active regions AR12565 and AR12567 gave us a confirmation of the method, though the MHD numerical code used in this work is only 2D one.

The complex formed by AR12565 and AR12567 acted many time as a quadrupolar unique region. The 3D coronal magnetic field extrapolations revealed that AR12567 appeared by flux emergence too close to AR12565, or even separated from this one from one of its side. In this way, their evolution can not be a surprise, many flares occurring in the middle zone between the two active regions.
So, the main conclusion of this paper is that we actually have a single one region with development in more stages: a period being a bipolar region and later becoming a quadrupolar one.
The numerical simulations results shown a continuous plasma and magnetic flux emergence during the AR evolution. These processes led to a myriad of each day flares and a few of small CMEs or mass ejecta.

%%%%%%%%%%%%%%%%%%%%%%%bibtex%%%%%%%%%%%%%%%%%%%%%%%%%%%%%%%%%%%%%%%%
%\bibliographystyle{IEEEtran}
%\bibliography{cdumitrache}

%%%%%%%%%%%%%%%%%%%%%%%%%%%%%%%%%%%%%%%%%%%%%%%%%%%%%%%%%%%%%%%%%%%%%%%%%%%%%%%
%\makeatletter
%\def\@biblabel#1{}
%\makeatother

\end{document}